\documentstyle[11pt,IAU207_pasp,twoside,psfig]{article}

\def\edcomment#1{\iffalse\marginpar{\raggedright\sl#1\/}\else\relax\fi}
\reversemarginpar

\begin{document}

\title{Formation and Disruption of Globular Star Clusters}

\author{S. Michael Fall}

\affil{Space Telescope Science Institute, 3700 San Martin Drive,
Baltimore, MD 21218, USA} 

\author{Qing Zhang}

\affil{Space Telescope Science Institute, 3700 San Martin Drive,
Baltimore, MD 21218, USA}

\begin{abstract}

In the first part of this article, we review observations of the 
mass and luminosity functions of young and old star cluster systems.
We also review some of the physical processes that may determine
the characteristic mass of globular clusters and the form of their 
mass function.
In the second part of this article, we summarize our models for the 
disruption of clusters and the corresponding evolution of the mass
function.
Much of our focus here is on understanding why the mass function
of globular clusters has no more than a weak dependence on radius 
within their host galaxies.

\end{abstract}

\section{Background}

The most basic attribute of any population of astronomical objects 
is its mass function.
In our notation, $\psi(M)$ is the number of objects per unit mass 
$M$, and $\Psi(\log M)$ is the number of objects per unit $\log M$. 
These two forms of the mass function are related by $\Psi(\log M) 
= (\log e)^{-1}M\psi(M)$.
They contains information about the physical processes involved in 
the formation and subsequent evolution of the objects.
There are well-known reasons, for example, why stars and galaxies
have roughly the masses they do and not some others, even if we 
lack definitive theories for the detailed forms of the stellar
and galactic mass functions.
This article addresses the question: What physical processes
determine the mass function of star clusters, especially that 
of globular clusters?

The upper panel of Figure~1 shows the empirical mass function 
of young star clusters in the interacting and merging Antennae 
galaxies (from Zhang \& Fall 1999).
This function declines monotonically, approximately as $\psi(M)
\propto M^{-2}$, over the entire observed range, $10^4\la M \la 
10^6~M_{\odot}$.
In a young cluster system, such as the one in the Antennae
galaxies, where the spread in the ages of the clusters is 
comparable with their median age, the luminosity function 
need not reflect the underlying mass function, since the
clusters have a wide range of mass-to-light ratios. 
However, luminosity functions are easier to determine than mass 
functions and are known for more cluster systems.
For all the young cluster systems studied so far, the luminosity
functions are well-approximated by power laws (Milky Way, van den 
Bergh \& Lafontaine 1984; LMC, Elson \& Fall 1985; M33, Christian 
\& Schommer 1988; Antennae, Whitmore et al. 1999).
In fact, the mass and luminosity functions of young star clusters
are remarkably similar to the mass function of interstellar clouds 
in the Milky Way (Dickey \& Garwood 1989; Solomon \& Rivolo 1989),
as emphasized by several authors (Harris \& Pudritz 1994;
Elmegreen \& Efremov 1997).

\begin{figure}
\centerline{\vbox{\psfig{figure=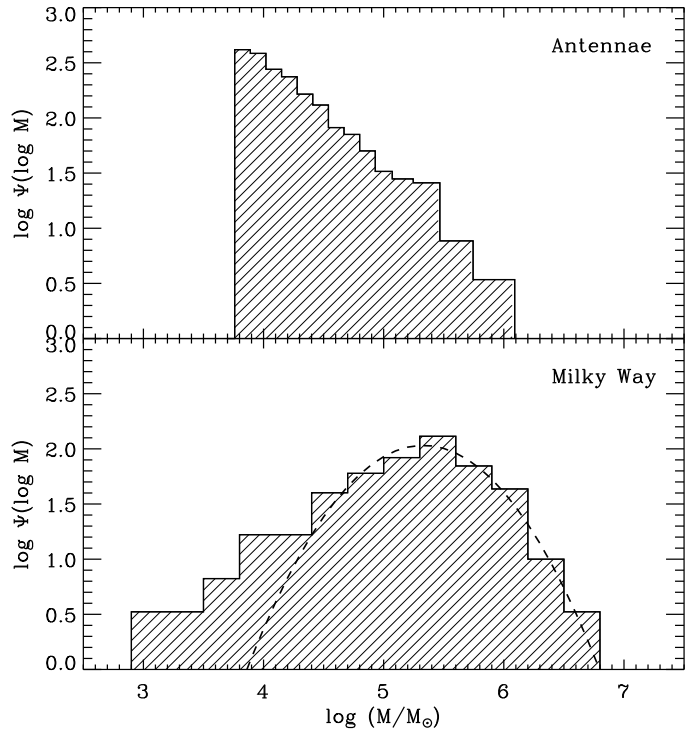,width=9cm,angle=0}}}
\caption{
Empirical mass functions of young star clusters in the Antennae 
galaxies and old globular clusters in the Milky Way. 
The former is from Zhang \& Fall (1999); the latter is based on data 
compiled by Harris (1996, 1999). 
The dashed curve is the usual lognormal representation of the mass 
function, corresponding to a Gaussian distribution of magnitudes.
}
\end{figure}

The lower panel of Figure~1 shows the empirical mass function old 
globular clusters in the Milky Way.
This was derived from the luminosities of all the clusters in the
Harris (1996, 1999) compilation, with a fixed mass-to-light ratio 
($M/L_V=3$), since the spread in the ages of the clusters is 
smaller than their median age.
The mass function of the globular clusters in the Milky Way, like
those in the spheroids of other well-studied galaxies, rises to 
a peak or turnover at $M_p\approx 2\times 10^5~M_{\odot}$ and then 
declines.
The corresponding feature in the luminosity function, at $\bar M_V 
\approx -7.3$, is sometimes used as a standard candle in distance 
determinations.
The empirical mass function is often represented by a lognormal 
function, although, as Figure~1 indicates, the former is actually
shallower than the latter for small masses.
The crucial point here is that the mass and luminosity functions
of young cluster systems are scale-free, whereas those of old
cluster systems have a preferred scale. Why is this?

Two explanations have been proposed for the age-dependence of 
the mass functions.
The first is that the conditions in ancient galaxies and 
protogalaxies favored the formation of clusters with masses 
$\sim 10^5$--$10^6~M_{\odot}$ but that these conditions no 
longer prevail in modern galaxies. 
For example, the Jeans mass could have been much higher in the
past, as a result of less efficient cooling by heavy elements 
and/or more efficient dissociation of molecular hydrogen (Fall 
\& Rees 1985; Kang et al. 1990).
It is sometimes stated in observational papers that this theory 
is ruled out by the lack of correlation between the luminosities
(masses) of globular clusters and their metallicities.
However, this argument is not correct, because the dependence of
the Jeans mass on the abundances of heavy elements and molecules 
is essentially bimodal.
The relevant Jeans mass is determined by whether or not gas in 
the protoclusters cools rapidly at temperatures below $10^4$~K.
If it does, the Jeans mass is $\la 10^2~M_{\odot}$; if it does 
not, the Jeans mass is $\sim 10^6~M_{\odot}$.
In the first case, one is making stellar associations or small 
open clusters; in the second, one is making globular clusters.

\begin{figure}
\centerline{\vbox{\psfig{figure=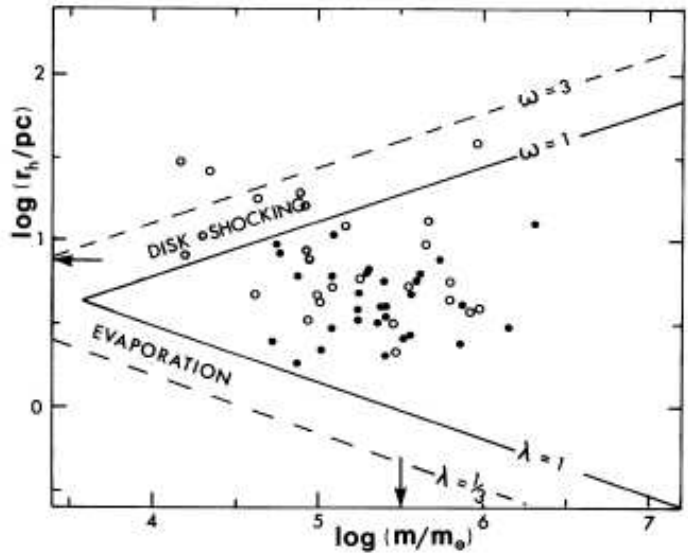,width=9cm,angle=270}}}
\caption{
Survival triangle in the mass-radius plane from Fall \& Rees (1977).
The solid lines show where the timescales for disruption by disk 
shocks and two-body relaxation are equal to the Hubble time.
The dashed lines indicate some of the uncertainties in these 
timescales.
The dots represent globular clusters in the Milky Way; filled and
open symbols indicate clusters closer to and farther from the 
Galactic center than the Sun (about 8~kpc).
Clusters outside the triangle will be destroyed within the next
Hubble time, whereas those inside will survive for longer. 
}
\end{figure}

The other explanation for the differences between the mass functions
of young and old cluster systems is that they represent the erosion
of the initial mass function by the gradual disruption of low-mass
clusters.
Star clusters are relatively weakly bound objects and are vulnerable
to disruption by a variety of processes, including mass loss by
stellar evolution (supernovae, stellar winds, etc.), evaporation 
by two-body relaxation and gravitational shocks (against the 
galactic disk and bulge), and dynamical friction (see Spitzer 
1987 for a review).
Figure~2 shows the survival region for globular clusters in the 
mass-radius plane defined by some of these processes (from Fall 
\& Rees 1977).
Recent work shows that disruptive processes, especially two-body 
relaxation, operating for a Hubble time, would cause the mass 
function to evolve from a variety of initial forms into one 
resembling that of old globular clusters (Vesperini 1997, 
1998; Baumgardt 1998; Fall \& Zhang 2001).

There is, however, a potentially serious objection to the idea
that disruption is responsible for the low-mass form of the mass 
function of old globular clusters:
the chief disruptive processes operate at different rates in 
different parts and different types of galaxies (Caputo \& 
Castellani 1984; Gnedin \& Ostriker 1997).
For example, the rate at which stars escape by two-body relaxation 
depends on the density of a cluster, which is determined by the 
tidal field, and hence is higher in the inner parts of galaxies 
than in the outer parts.
The rate at which stars escape by gravitational shocks is also 
higher in the inner parts of galaxies because the orbital periods 
are shorter and the surface density of the disk is higher there. 
Moreover, disks are absent in elliptical galaxies.
Thus, if the mass function were strongly affected by disruptive 
processes, one might expect its form to depend on radius within a 
galaxy and to vary from one galaxy to another. 
This, however, is contradicted by many observations showing that 
the mass function of old globular clusters varies little, if at 
all, within and among galaxies (Harris 1991). 

\section{New Models}

With these issues in mind, we have developed some new models
to compute the evolution of the mass function of star cluster
systems (Fall \& Zhang 2001).
Our models include disruption by two-body relaxation, gravitational 
shocks, and stellar evolution.
We describe these processes by approximate formulae that can be 
solved largely analytically. 
The clusters are assumed to orbit in a galactic potential that is
static, spherical, and has a logarithmic dependence on the distance
$R$ from the galactic center. 
Each cluster is assume to be tidally limited at the pericenter
of its orbit and to lose mass at a constant mean internal density. 
The population of orbits is specified by the distribution function
$f(E,J)$, defined as the number of clusters per unit volume of
position-velocity space with energy and angular momentum per
unit mass near $E$ and $J$.
The distribution function determines how much the clusters are 
mixed in radius and hence how much the mass function varies with 
radius.

We consider two simple models for the initial distribution function.
The first is the Eddington model
\begin{equation}
f_0(E,J) \propto \exp(-E/\sigma^2) \exp[-\onehalf (J/R_A \sigma)^2].
\end{equation}
This has velocity dispersions $\sigma_R=\sigma$ and 
$\sigma_T=\sigma[1+(R/R_A)^2]^{-1/2}$ in the radial and transverse 
directions, where the anisotropy radius $R_A$ marks the transition 
from a nearly isotropic to a predominantly radial velocity distribution.
The second initial distribution function we consider has the form 
\begin{equation}
f_0(E,J) \propto \exp(-E/\sigma^2)  J^{-2\beta}.
\end{equation}
In this case, the radial and transverse velocity dispersions are
$\sigma_R=\sigma$ and $\sigma_T=\sigma(1-\beta)^{1/2}$.
We refer to this as the scale-free model. 
In most cases, we adopt $R_A = 5$~kpc and $\beta = 0.5$, so that 
both models have the same velocity anisotropy at the median radius 
of the globular cluster system ($R_h = 5$~kpc).
For our purposes, the most important difference between the 
Eddington and scale-free models is that, in the former, the 
velocity anisotropy increases outward, whereas in the latter, 
it is the same at all radii. 
Thus, the distribution of pericenters is narrower in the 
Eddington model than it is in the scale-free model, as shown
in Figure~3.

\begin{figure}
\centerline{\vbox{\psfig{figure=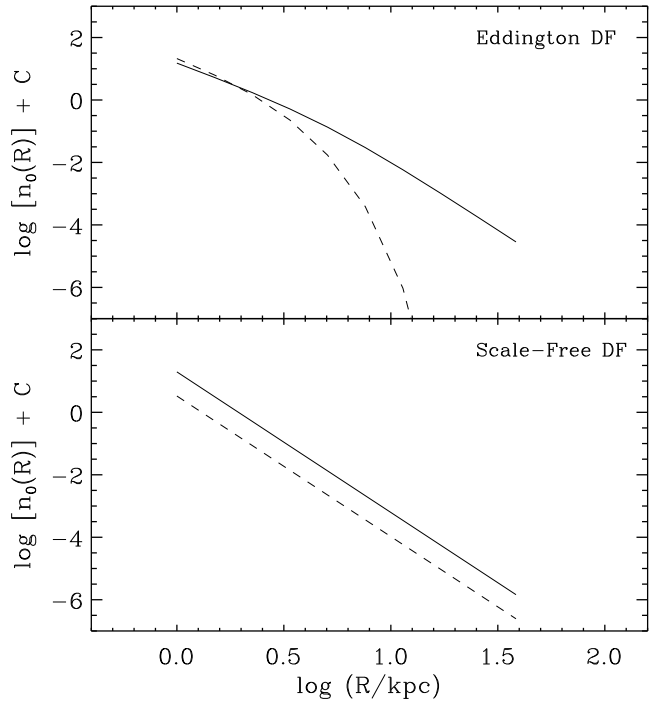,width=8.5cm,angle=0}}}
\caption{  
Initial densities of clusters positions (solid lines) and 
pericenters (dashed lines) for the Eddington and scale-free 
distribution functions. 
Note that the distribution of pericenters is narrower than the
distribution of cluster positions for the Eddington model but
not for the scale-free model.
}
\end{figure}

\begin{figure}
\centerline{\vbox{\psfig{figure=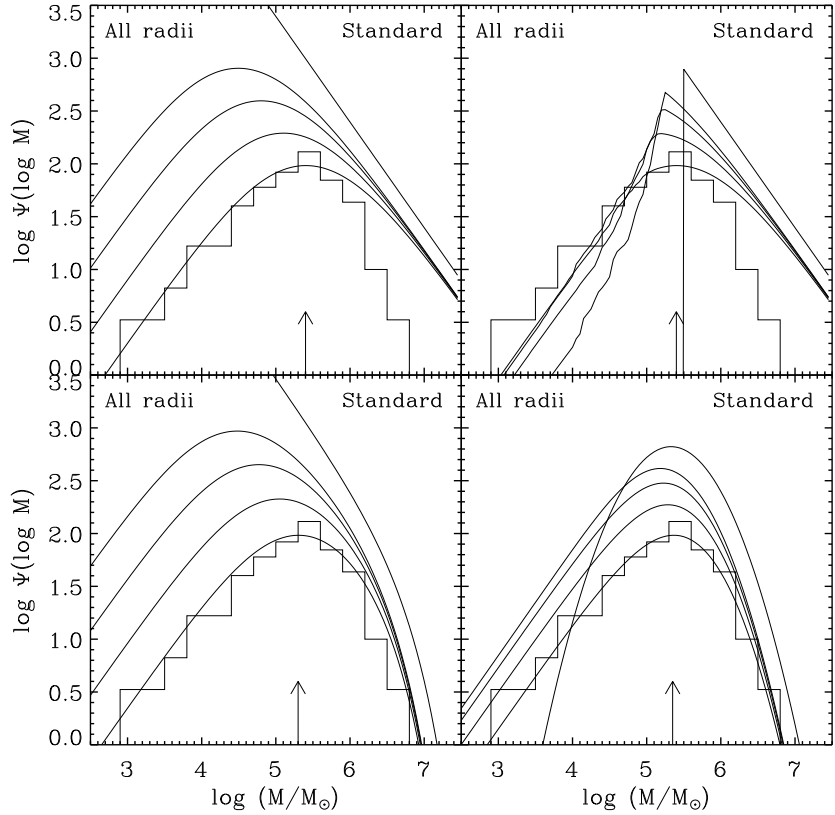,width=11cm,angle=90}}}
\caption{  
Evolution of the mass function, averaged over all radii, for the
Eddington initial distribution function and four different initial 
mass functions. 
These are (clockwise from upper left): a pure power law, a truncated 
power law, a Schechter function, and a lognormal function.
Each mass function is plotted at $t=0$, 1.5, 3, 6, and 12~Gyr;
the arrows indicate the peak at $t=12$~Gyr. 
The histograms depict the empirical mass function of globular 
clusters in the Milky Way (the same as in Fig.~1). 
Note that the peak mass in the models is similar to that in the 
observations for the four different initial conditions.
}
\end{figure}

We consider four models for the initial mass function of the
clusters: (1) a pure power law, $\psi_0(M)\propto M^{-2}$, (2) the
same power law truncated at $M_l = 3\times10^5~M_{\odot}$, (3) a
Schechter function, and (4) a lognormal function.
Figure~4 shows the evolution of the mass function, averaged over 
all radii, for the Eddington initial distribution function.
In all four cases, the mass function develops a peak, which, after
12~Gyr is remarkably close to the observed peak, despite the very
different initial conditions.
Below the peak, the evolution is dominated by two-body relaxation,
and the mass function always develops a low-mass tail of the form 
$\psi(M) = {\rm const}$. 
This can be traced to the fact that, in the late stages of disruption,
the masses of tidally limited clusters decrease linearly with time.
The predicted low-mass form of the mass function agrees nicely with 
the observed form.
Above the peak, the evolution of the mass function is dominated 
by stellar evolution at early times and by gravitational shocks 
at late times. 
These processes shift the mass function to lower masses but leave 
its shape nearly invariant.
Thus, the present shape of the mass function at high masses is 
largely determined by its initial shape.
The radially averaged mass function for the scale-free model (not 
shown) is similar to that for the Eddington model.
The evolution of the density profiles of the cluster systems for 
both models are shown in Figure~5.

\begin{figure}
\centerline{\vbox{\psfig{figure=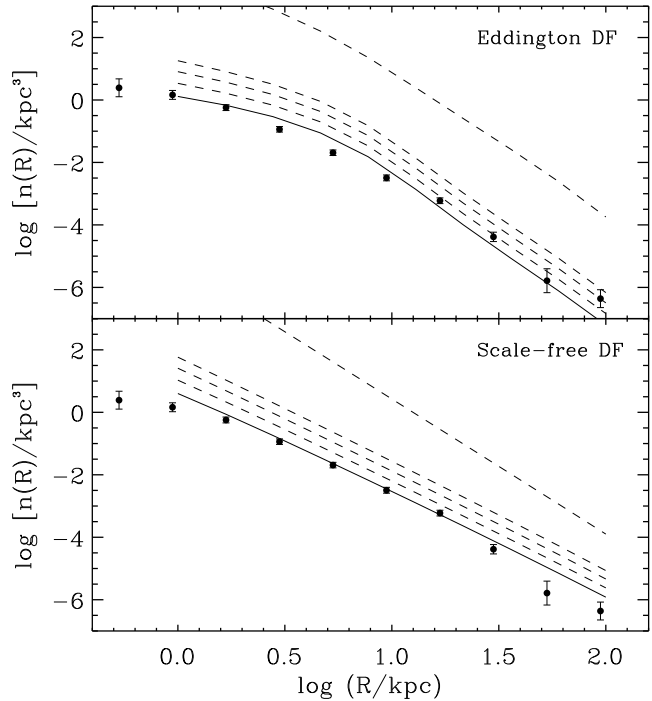,width=8.5cm,angle=0}}}
\caption{  
Evolution of the number density profile of the cluster system for
the Eddington and scale-free initial distribution functions.
The profiles are plotted at $t=0$, 1.5, 3, 6, and 12~Gyr. 
The data points depict the empirical profile for globular clusters
in the Milky Way.
Note that the final profiles in the models are in reasonable 
agreement with the empirical profile.
}
\end{figure}

\begin{figure}
\centerline{\vbox{\psfig{figure=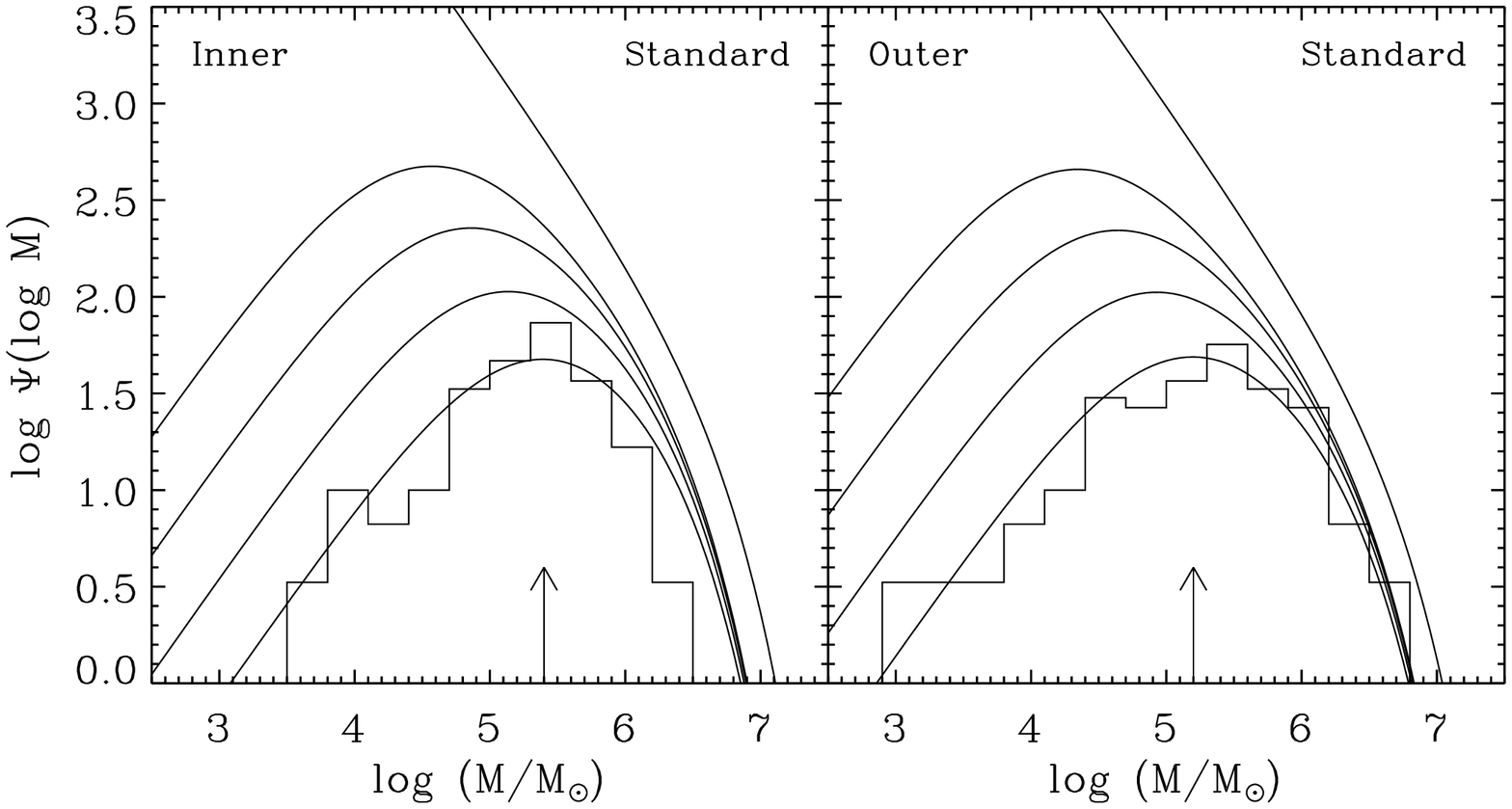,width=11.5cm,angle=90}}}
\caption{  
Evolution of the mass function, averaged over inner radii ($R < 5$~kpc)
and outer radii ($R > 5$~kpc), for the Eddington initial distribution 
function and the Schechter initial mass function.
Each mass function is plotted at $t=0$, 1.5, 3, 6, and 12~Gyr;
the arrows indicate the peak at $t=12$~Gyr. 
The histograms depict the empirical mass functions of globular 
clusters in the Milky Way in the corresponding ranges of radii. 
Note that the shift in the peak mass in the models between inner 
and outer radii is relatively small.
}
\end{figure}

\begin{figure}
\centerline{\vbox{\psfig{figure=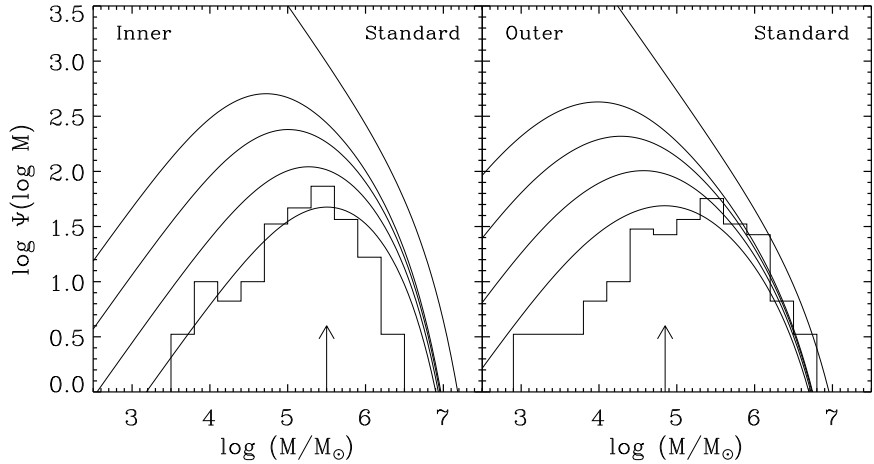,width=11.5cm,angle=90}}}
\caption{  
Evolution of the mass function, averaged over inner radii ($R < 5$~kpc)
and outer radii ($R > 5$~kpc), for the scale-free initial distribution 
and the Schechter initial mass function.
Each mass function is plotted at $t=0$, 1.5, 3, 6, and 12~Gyr;
the arrows indicate the peak at $t=12$~Gyr. 
The histograms depict the empirical mass functions of globular 
clusters in the Milky Way in the corresponding ranges of radii. 
Note that the shift in the peak mass in the models between inner 
and outer radii is relatively large.
}
\end{figure}

Where the Eddington and scale-free models differ most is in 
the radial variation of the mass function of the clusters.
This is shown in Figures~6 and~7, which display the evolution 
separately for clusters inside and outside $R=5$~kpc.
In both models, the peak mass is larger at small radii.
This is caused mainly by the higher rate of evaporation by 
two-body relaxation, resulting from the larger mean densities 
of clusters with small pericenter distances. 
In the Eddington model, the radial variation of the peak mass 
is weak enough to be consistent with observations, whereas in 
the scale-free model, the variation is too strong.
The reason for this is that the distribution of pericenters
is narrower in the Eddington model, leading to a smaller
range of disruption rates, than in the scale-free model.

The lesson here is that radial mixing is a necessary but not
sufficient condition for weak radial variations in the mass
function of globular clusters.
The other requirement is that the radial anisotropy in the 
initial velocity distribution of the clusters increase outward, 
as in the Eddington model.
The present velocity distribution of globular clusters in the Milky 
Way appears to have little or no radial anisotropy (Frenk \& White 
1980; Dinescu, Girard, \& van Altena 1999).
This is qualitatively what we would expect, since most clusters on 
elongated orbits would already have been destroyed, leaving behind 
a nearly isotropic or tangentially biased velocity distribution.
These conclusions are based on models with static, spherical
galactic potentials, in which each cluster returns to the same
pericenter on each of its revolutions around a galaxy.
In galaxies with time-dependent and/or non-spherical potentials,
however, the pericenters of the clusters may change from one 
revolution to the next.
This will tend to homogenize the disruption rates of the clusters
and hence to weaken the radial variation in their mass function.
Whether these effects are significant in galaxies like the Milky 
Way is not yet known.

It is important to test the models of disruption against observations.
A clear prediction of the models is that the peak mass $M_p$ should 
increase with the ages of clusters. 
This might be observable in galaxies in which clusters formed 
continuously over long periods of time. 
Alternatively, the evolution of the peak mass might be observable 
in galaxies with bursts of cluster formation at different times,
such as in a sequence of merger remnants. 
This test may be difficult, however, because the luminosity 
corresponding to the peak mass is relatively small for young clusters 
(since $M_p$ varies more rapidly with $t$ than $M/L_V$ does).
Another prediction of the models is that the peak mass should decrease 
with increasing distance from the centers of galaxies, unless this has 
been completely diluted by the mixing of pericenters mentioned above.
Searches for radial variations in the peak mass have so far been 
inconclusive. 
This test is difficult because the diffuse light of the host galaxies 
also varies with radius, making it harder to find faint clusters in 
the inner regions. 
Finally, the strong dependence of the peak mass on the ages of clusters
and the weak dependence on their positions within and among galaxies 
cast some doubt on the use of the peak luminosity as a standard candle 
for distance estimates. 
This method may be viable, however, if the samples of clusters are
carefully chosen from similar locations in similar galaxies.

The models of disruption help to dissolve some of the boundaries
in the classification of star clusters.
The shape of the mass function above the peak is largely preserved 
as clusters are disrupted and hence should reflect processes at the 
time they formed.
Below the peak, however, the shape of the mass function is determined
entirely by disruption, mainly driven by two-body relaxation, and hence 
contains no information about how the clusters formed. 
If there were any feature in the initial mass function, such as a 
Jeans-type lower cutoff, it would have been erased.
In our models, the only feature in the present mass function, the peak 
at $2\times 10^5~M_{\odot}$, is largely determined by the condition that 
clusters of this mass have a timescale for disruption comparable to 
the Hubble time. 
Thus, it is conceivable that star clusters of different types (open, 
populous, globular, etc.) formed by the same physical processes with 
the same initial mass function and that the differences in their 
present mass functions reflect only their different ages and local 
environments, primarily the strength of the galactic tidal field. 
Our results therefore support the suggestion that at least some of 
the star clusters formed in merging and other starburst galaxies may 
be regarded as young globular clusters. 
Further investigations of these objects may shed light on the 
processes by which old globular clusters formed.


\begin{references}

\reference Baumgardt, H. 1998, \aap, 330, 480
\reference Caputo, F., \& Castellani, V. 1984, \mnras, 207, 185
\reference Christian, C. A., \& Schommer, R. A. 1988, \aj, 95, 704
\reference Dickey, J. M., \& Garwood, R. W. 1989, \apj, 341, 201
\reference Dinescu, D. I., Girard, T. M., \& van Altena, W. F. 1999, 
           \aj, 117, 1792
\reference Elmegreen, B. G., \& Efremov, Y. N. 1997, ApJ, 480, 235
\reference Elson, R. A. W., \& Fall, S. M. 1985, \pasp, 97, 692
\reference Fall, S. M., \& Rees, M. J. 1977, \mnras, 181, 37P
\reference Fall, S. M., \& Rees, M. J. 1985, \apj, 298, 18
\reference Fall, S. M., \& Zhang, Q. 2001, \apj, in press 
           (astro-ph/0107298)
\reference Frenk, C. S., \& White, S. D. M. 1980, \mnras, 193, 295
\reference Gnedin, O. Y., \& Ostriker, J. P. 1997, \apj, 474, 223
\reference Harris, W. E. 1991, \araa, 29, 543
\reference Harris, W. E. 1996, \aj, 112, 1487
\reference Harris, W. E. 1999, http://www.physics.mcmaster.ca/Globular.html
\reference Harris, W. E., \& Pudritz, R. E. 1994, ApJ, 429, 177
\reference Kang, H., Shapiro, P. R., Fall, S. M., \& Rees, M. J. 1990,
           \apj, 363, 488
\reference Solomon, P. M., \& Rivolo, A. R. 1989, \apj, 339, 919
\reference Spitzer, L. 1987, Dynamical Evolution of Globular Clusters 
           (Princeton: Princeton Univ. Press)
\reference van den Bergh, S., \& Lafontaine, A. 1984, \aj, 89, 1822
\reference Vesperini, E. 1997, \mnras, 287, 915
\reference Vesperini, E. 1998, \mnras, 299, 1019
\reference Whitmore, B. C., Zhang, Q., Leitherer, C., Fall, S. M., 
           Schweizer, F., \& Miller, B. W. 1999, \aj, 118, 1551
\reference Zhang, Q., \& Fall, S. M. 1999, \apj, 527, L81

\end{references}
\end{document}